\documentclass[lettersize,journal]{IEEEtran}
\usepackage{amsmath,amsfonts}
\usepackage{algorithmic}
\usepackage{algorithm}
\usepackage{array}
\usepackage[caption=false,font=normalsize,labelfont=sf,textfont=sf]{subfig}
\usepackage{nomencl}
\usepackage{textcomp}
\usepackage{stfloats}
\usepackage{url}
\usepackage{verbatim}
\usepackage{graphicx}
\usepackage{cite}
\usepackage{todonotes}
\makenomenclature

\everymath=\expandafter{\the\everymath\displaystyle}

\usepackage{siunitx}
\begin{document}

\nomenclature{$\Phi$}{Cost function}
\nomenclature{$y$}{Measurement vector}
\nomenclature{$Y$}{Output constraint set}
\nomenclature{$u$}{Set point vector}
\nomenclature{$U$}{Input constraint set}
\nomenclature{$h$}{Steady-state input to output map}
\nomenclature{$A$}{Input constraint matrix}
\nomenclature{$C$}{Output constraint matrix}
\nomenclature{$b$}{Input constraint vector}
\nomenclature{$d$}{Output constraint vector}
\nomenclature{$p$}{Number of set points}
\nomenclature{$n$}{Number of measurements}
\nomenclature{$q$}{Number of input constraints}
\nomenclature{$l$}{Number of output constraints}
\nomenclature{$\Delta h(u)$}{Sensitivity matrix}
\nomenclature{$\alpha$}{Controller gain}
\nomenclature{$\hat{\sigma}_{\alpha}$}{Optimization result, optimal step}
\nomenclature{$w$}{Optimization variable}
\nomenclature{$F$}{Set of controllable devices}
\nomenclature{$B$}{Set of observed lines}
\nomenclature{$N$}{Set of observed buses}
\nomenclature{$P_{set}$}{Active power set point}
\nomenclature{$P_{planned}$}{Original active power set point}
\nomenclature{$P_{meas}$}{Measured active power}
\nomenclature{$Q_{set}$}{Reactive power set point}
\nomenclature{$P_{max/min}$}{Maximum/Minimum active power}
\nomenclature{$Q_{max/min}$}{Maximum/Minimum reactive power}
\nomenclature{$V_{meas}$}{Measured bus voltage}
\nomenclature{$I_{meas}$}{Measured line current}
\nomenclature{$V_{max/min}$}{Maximum/Minimum bus voltage}
\nomenclature{$I_{max/min}$}{Maximum/Minimum line current}

%\title{Tuning a Cascaded Online Feedback Optimization Controller for Provision of Distributed Flexibility \\
%}
\title{Influence of Controller Tuning on Cascaded Flexibility Provision with Feedback Optimization \\
}

\author{Irina Zettl,~\IEEEmembership{Student Member,~IEEE}, Florian Klein-Helmkamp,~\IEEEmembership{Student Member,~IEEE}, Florian Schmidtke, Lukas Ortmann, Andreas Ulbig~\IEEEmembership{Senior Member,~IEEE}
        % <-this % stops a space
\thanks{\copyright 2026 IEEE.  Personal use of this material is permitted.  Permission from IEEE must be obtained for all other uses, in any current or future media, including reprinting/republishing this material for advertising or promotional purposes, creating new collective works, for resale or redistribution to servers or lists, or reuse of any copyrighted component of this work in other works.}% <-this % stops a space
%\thanks{Manuscript received April 19, 2021; revised August 16, 2021.}
}

\markboth{IEEE TRANSACTIONS ON INDUSTRY APPLICATIONS}%
{Zettl \MakeLowercase{\textit{et al.}}: Tuning Online Feedback Controllers for Distributed Flexibility Provision}

\IEEEpubid{0000--0000/00\$31.00~\copyright~2026 IEEE}

\maketitle

\begin{abstract}
The coordination of a large number of flexibility-providing units across various grid layers requires innovative control concepts. This is needed, e.g., to allow active distribution systems to provide ancillary services for the transmission system. A cascaded control structure based on Online Feedback Optimization (OFO) can be used to meet flexibility requests at the point of common coupling by tracking an active power set point at the point of common coupling. This paper investigates the practical influence of the parameterization of the individual controllers on the performance of the hierarchical flexibility provision in three case studies. One case study includes a two-level controller cascade acting on one medium and two low voltage grids, and the other one includes a three-level cascade acting on low to high voltage levels. The results show that the behavior of one controller is highly dependent on the choice of control parameters of the other controllers in the cascade. Additionally, the choice of parametrization has a significant impact on the accuracy and speed of flexibility provision. A third case study investigates the effects of model mismatch and measurement noise on the appropriate selection of parameters.  Overall, careful tuning enables the efficient vertical coordination of flexibility-providing units with a cascaded structure based on OFO.
 \\
\end{abstract}

\begin{IEEEkeywords}
ancillary services, controller tuning, distributed flexibility, online feedback optimization.
\end{IEEEkeywords}

\vspace{-3mm}
\printnomenclature[1.75cm]

\IEEEpubidadjcol
\section{Introduction}
\label{sec:intro}
\IEEEPARstart{H}{igher} fluctuations in the utilization of the grid infrastructure and assets, and the occurrence of bidirectional power flows, are two consequences of the replacement of large conventional power plants with small-scale, volatile generation units~\cite{bmwk2023}. 
A cost-effective alternative to grid expansion and reinforcement is a higher utilization of existing grid infrastructure~\cite{innosys2030}. To realize this, certain ancillary services, e.g. curative congestion management, can be used.  
Active distribution grids can provide flexibility for such ancillary services by actuating the small-scale distributed energy resources (DER) connected to them. Together with the increasing spatial distribution of DERs and their connection to multiple voltage levels, this is driving research into new control and coordination concepts. In particular, this includes the integration of distributed flexibility into transmission system operation, e.g. through preventive redispatch~\cite{hoffrichter2019} or curative measures~\cite{graupner2022}. These developments give rise to the so-called vertical grid operation~\cite{frueh2023}, in which flexibility from underlying grid levels is abstracted and made available at system interfaces for use by superordinate entities~\cite{kolster2020}.

Within the transmission or distribution system, flexibility providing units (FPUs) can be connected to different voltage levels depending on their power rating. Utilizing the flexibility provided by these FPUs requires coordination across multiple grid layers. For this, an exchange of information and therefore communication between different system operators is needed. For this type of problem, solutions including an optimal power flow (OPF) are used in many cases \cite{contreras2021}. Approaches like these rely on accurate grid models, although they are not always available. Vertical coordination of multiple FPUs with limited need for information exchange between system operators can be achieved by using online feedback optimization (OFO) within a cascaded control architecture. OFO tracks the optimal solution of the dispatch problem by integrating an optimization algorithm in closed loop with the physical grid. Considering the actual grid state by integrating real-time measurements in the optimization with the feedback structure of the controller makes the approach robust to disturbances and inaccurate grid models \cite{klein-helmkamp2024}.

\begin{figure}[tb]
	\centerline{\includegraphics{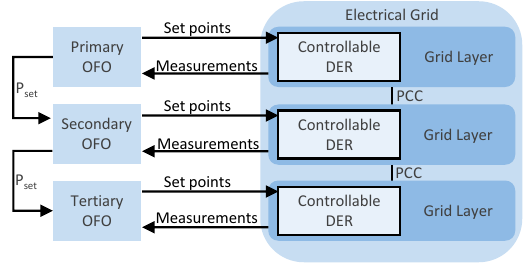}}
	\caption{Cascaded structure of OFO controllers and grid layers \cite{zettl2024}}
	\label{fig:ofo_simple}
\end{figure}

A cascaded structure incorporating one or more OFO controllers for each grid layer can be used to control the active power flow at the point of common coupling (PCC) with the overlaying grid  (see Fig.~\ref{fig:ofo_simple}) and fulfill a flexibility request. With this, a bottom-up aggregation of flexibility is achieved.
The advantages of this algorithm can thus be used for efficient vertical coordination of FPUs. The parametrization of individual controllers can affect the performance of the entire cascade and therefore influence the flexibility provision. Hence, a careful tuning of the entire cascade is necessary.

\subsection{Related Work}
\label{sec:relatedwork}
Active distribution grids provide valuable flexibility, but careful coordination between various system operators and the respective DERs is needed  \cite{kolster2022}. In \cite{frueh2023} and \cite{kalantar2021}, among others, feedforward optimization-based coordination algorithms have been proposed. Employing online optimization integrated in feedback control schemes tackles the need for an accurate system model, a lack of robustness against model mismatch and other drawbacks of feedforward methods. \\
Previous work can be found concerning the stability and robustness of feedback optimization by exploring its theoretical foundations. In \cite{molzahn2017}, a survey on distributed optimization and control algorithms is performed, mentioning also online algorithms solving optimization and control problems in power systems. The survey in \cite{krishnamoorty2022} focuses on online optimization algorithms in feedback control schemes. An overview of different optimization algorithms used in feedback optimizations is given in \cite{hauswirth2024}, in addition to mechanisms ensuring closed-loop stability.  While \cite{colombino2020} used an augmented Lagrangian primal-dual saddle-flow algorithm in closed loop with the system, \cite{haberle2020} proposed a feedback controller applying a projected gradient approach. A model-free feedback controller is proposed in \cite{he2020}, updating the control inputs using estimates of the gradient.  

\cite{colombino2019} proposes a framework to asses the stability of online feedback optimization algorithms in a systematic way. A comprehensive robustness assessment for OFO controllers in power systems using primal-dual gradient projections is presented in \cite{Zhen_2025}. Stability of OFO schemes based on projected gradient dynamics has been established under explicit time-scale separation between the optimization loop and the physical plant. For LTI systems, computable upper bounds on the feedback gain guaranteeing closed-loop stability and convergence to optimal steady states are derived in \cite{menta2018}, with the bounds depending on plant stability margins and Lipschitz constants of the (projected) cost gradient. These results are generalized to projected gradient flows interconnected with asymptotically stable plants using
singular perturbation arguments in \cite{hauswirth2021}. From a complementary robust-control viewpoint, interpreting optimization algorithms as low-gain feedback controllers yields small-gain and passivity-based stability guarantees, providing theoretical justification for conservative tuning choices in OFO implementations \cite{hauswirth2024}.

While this paper focuses on the practical implications of controller tuning, a method to compute safe trajectory sets and analyze the controller behavior on the two-dimensional PQ-plane is proposed in \cite{klein-helmkamp2025a}.
Apart from the theoretical work, research was also conducted on  utilization in the context of power system operation and flexibility provision from distributed resources. 
In \cite{hauswirth2017}, OFO is used in a simulative study to steer the power system in real time to the optimal operating point without solving the AC OPF problems explicitly.
\cite{picallo2022} uses an OFO approach to track the solution of a time-varying OPF and proposes recursive least-squares estimation, basically turning the approach model-free. In \cite{ortmann2024a}, an OFO controller is analyzed using a benchmark subtransmission grid, showing its robustness against model mismatch. \cite{Zhan_2024} proposes a combination of localized DER control with OFO to prevent limit violations in real-time operation for unbalanced low voltage feeders. While most of the publications assume noiseless and full measurements, \cite{picallo2020} proposes a combination of OFO with a dynamic State Estimation. \cite{ortmann2023} specifically implements an OFO controller for congestion management using curative actions. 

Complementary to the theoretical considerations mentioned above, experimental validations of the OFO control scheme in power systems operation have been carried out: \cite{ortmann2020} compared the dispatch for voltage control with OFO to a dispatch with OPF, and \cite{ortmann2023b} implemented OFO in a real distribution grid. A laboratory setup was used in \cite{klein-helmkamp2023} to show the ability of OFO to coordinate multiple DERs for flexibility provision in an experimental validation. The general cascaded structure enabling coordination across various grid layers and system operators for flexibility provision has been presented in \cite{klein-helmkamp2024}. However, this has been done without further investigating the tuning of individual controllers. Further research on the tuning of OFO controllers for curative system operation was conducted in \cite{ortmann2024b}, and for set-point tracking in \cite{zagorowska2024}. However, the effects in a cascaded structure were not studied. Considering the system's full operating range for parameter choice when using a cascaded structure can be relevant for efficient operation \cite{Klein-Helmkamp2025}.

\subsection{Main Contribution}
This paper implements a cascaded control structure based on OFO with the aim of coordinating a large number of FPUs across multiple grid layers and different system operators and investigates the tuning thereof. The cascaded structure contains multiple OFO controllers, and each grid layer is coordinated by one specific controller within the cascade. One level of the cascade can contain multiple OFO controllers, but every controller observes and controls its own area of the grid. This architecture enables the top-down disaggregation of a flexibility request while aggregating the available flexibility bottom-up. 
Efficient tuning of individual controllers is necessary for fast and reliable control of active power flow at the PCC between different grid layers. Changing the controller gain leads to a change in convergence behavior of the respective controller, which in turn might influence the convergence behavior of the entire cascade. To investigate how this affects the flexibility provision to the overlaying grid layer, we formulate the following research questions: 

\begin{enumerate}
	\item How are the accuracy and speed of flexibility provision at the top of the cascade influenced by the choice of gain for an individual controller?
	\item How is the convergence to a steady state for OFO controllers in lower levels of the cascade affected by a changing flexibility request? 
    \item Which influence do model mismatch and measurement noise have on the optimal parameter choice?
\end{enumerate}
We implement three case studies, repeating and building on the tests performed in \cite{zettl2024}, to investigate the tuning of the cascade in different grid environments. The first case study is performed on a system with three levels being controlled by a single controller per level, while the second case study is performed on a two-layer system with multiple controllers receiving set points from the same superordinate controller.    
We expect that an increasing gain of individual controllers reduces the time needed for flexibility provision. Controllers might not be able to converge to a steady-state if other controllers do not converge as well due to suboptimal choice of parameters. The third case study gives recommendations for the selection of gain values in different scenarios. 

The remainder of the paper is structured in the following way: Section~\ref{ofointro} introduces OFO with a focus on the general optimization problem and the feedback nature of the approach. The implemented cascaded structure is then presented in Section~\ref{structure}. Finally, Sections~\ref{sec:case1}-\ref{sec:case3} provide the numerical results of the three performed simulative case studies.

\section{Online Feedback Optimization}
\label{ofointro}
The term online feedback optimization describes a control scheme consisting of an optimization problem and its solution that are incorporated in closed loop with online measurements from a physical system (e.g., the electrical power system). This creates a feedback control loop, as shown in Fig.~\ref{fig:ofo_complete}. In this work, we used OFO to control active power flow at the PCC between two grid layers while ensuring the satisfaction of grid constraints in the lower grid layer. This can be used to fulfill a flexibility request from the superordinate grid layer for the provision of different ancillary services. In this work, we focus on the control of active power flow for congestion management. The chosen formulation of the optimization problem, which is then solved with OFO during one iteration of the control algorithm, is presented in this section, and the feedback structure is explained. A hierarchical or cascaded control architecture was chosen, since the electricity grid consists of multiple voltage levels. The resulting hierarchy of controllers is explained in Section~\ref{structure}.

\subsection{Optimization Problem}
The overarching constrained optimization problem is given in (\ref{eq:genericOP}). The cost function, denoted as $\Phi$, can be freely chosen depending on the objective and could, e.g., be of techno-economic nature. To solve this optimization problem, (\ref{eq:qp}) is integrated into the OFO controller. From the solution, we receive the set point vector $u$, which serves as an input to the physical system. Vector $y$ is the output vector of the system, consisting of the measurements thereof. A nonlinear steady-state input-to-output map $y=h(u)$ links these two vectors.

\begin{equation}
	\label{eq:genericOP}
	\begin{aligned}
		\min_{u} \enspace & \Phi(u,y)  \\
		\text{s.t.} \enspace & u \in U \\
		& y \in Y \\
		& y = h(u) \\
		\text{with} \enspace & U = \{u \in \mathbb{R}^p | Au \leq b\} \\
		\text{and} \enspace &  Y = \{y \in \mathbb{R}^n | Cy \leq d\} \\
		\text{where} \enspace & A \in \mathbb{R}^{q \times p}, \enspace b \in \mathbb{R}^{q}, \enspace C \in \mathbb{R}^{l \times n} \enspace and \enspace d \in \mathbb{R}^{l}
	\end{aligned}
\end{equation}
In this work, a projected gradient descent scheme is used to solve (\ref{eq:genericOP}), which is possible as long as the cost function is continuously differentiable on $\mathbb{R}$. The gradient of the function is therefore updated iteratively. This leads to convergence to a minimizer of \ref{eq:genericOP} by proceeding one step in every iteration in the direction of the total derivative of the function. Since $h$ is continuously differentiable, the total derivative can be calculated as follows:

\begin{equation}
	\label{eq:totalderivative}
	\begin{aligned}
		\frac{d\Phi(u,y)}{du}  = & \nabla_u \Phi(u,y) \big|_{y=h(u)} \\ &  + \nabla h(u)^T \nabla_y \Phi(u,y) \big|_{y=h(u)}  \\
		 = & H(u)^T \nabla \Phi(u,y) \big|_{y=h(u)}
	\end{aligned}
\end{equation}
One advantage is that OFO does not require a detailed model of the system. The only explicit information about the system model that is needed is the sensitivity matrix $\nabla h(u)$. The matrix consists of steady-state sensitivities describing the influence of a change in the set point vector $u$ on the output vector $y$. It therefore represents a linearization of the power flow equations at an initial state $y_0$. To calculate highly accurate sensitivities, one would need complete information about the grid model and state. \cite{ortmann2020} has experimentally proven that using constant approximate sensitivities instead yields good results as well. We therefore calculate entries of the sensitivity matrix a-priori as follows:

\begin{equation}
	\nabla h_{i,j} = \frac{\partial h(u)_{i}}{\partial u_{j}}
\end{equation}

A quadratic program (QP) is solved to project the calculated gradient onto the set of feasible points. The QP is defined in (\ref{eq:qp}). $\hat{\sigma}_{\alpha}$ then denotes the best step that complies with the constraints while following  the direction of the total derivative.

\begin{equation}
	\label{eq:qp}
	\begin{aligned}
		\hat{\sigma}_{\alpha}(u,y) & = \arg \min_{w \in \mathbb{R}^p} ||w + H(u)^T \nabla \Phi(u,y) ||^2 \\
		\text{s.t.} \enspace &A(u+\alpha w) \leq b \\
		& C (y + \alpha \nabla h (u)w) \leq d \\
		\text{with} \enspace & H(u)^T = [ \mathbb{I}_p \nabla h(u)^T] \\
	\end{aligned}
\end{equation}

\subsection{Feedback Control}
\begin{figure}[tb]
	\centerline{\includegraphics{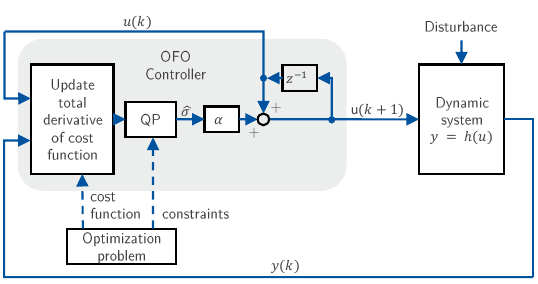}}
	\caption{OFO controller described as block diagram \cite{zettl2024}}
	\label{fig:ofo_complete}
\end{figure}

The optimization problem presented before is now integrated into a feedback control loop, as shown in Fig.~\ref{fig:ofo_complete}.
To iteratively determine the optimal system inputs, the following steps are performed during one iteration of the OFO algorithm:
\begin{enumerate}
	\item 	Receive updated measurement $y$ from physical system
	\item   Calculate new gradient$\nabla \Phi(u,y)$
	\item 	Determine optimal next step by solving QP
	\item 	Update set point vector $u_{t+1}$
	\item   Send new system inputs to the physical system
\end{enumerate} 	
The new system inputs $u_{t+1}$ are calculated similarly to the differential equation of a PI controller: The calculated step $\hat{\sigma}_{\alpha}$, scaled with the controller gain $\alpha \in \mathbb{R}_{>0}$, is added to the last known set point vector:

\begin{equation}
	\label{eq:nextinput}
	u_{k+1} = u + \alpha \hat{\sigma}_{\alpha}
\end{equation}
This step is performed iteratively, achieving asymptotic convergence to the optimal set point vector while enforcing the constraints regarding the system outputs. 
Compared to classical feedforward approaches, the feedback nature of OFO holds some advantages. As discussed, a detailed system model is not needed; hence, OFO presents a suitable alternative, e.g. to the use of OPF, for all cases where explicit model information of the controlled grid beyond the sensitivity matrix is not available or inaccurate. It takes the current operating point of the system into account by considering real-time measurements, and it displays robustness against model mismatch. Constraint satisfaction is ensured even with inaccurate information. OFO can be used as a real-time controller, as low computational complexity allows fast satisfaction of grid constraints \cite{ortmann2023}. In this work, we assume timescale separation between the control actions and fast system dynamics by assuming that an adequate cycle time is chosen \cite{hauswirth2021}.

\section{Hierarchical Flexibility Coordination}
\label{structure}
A multitude of DERs connected to different grid layers in the distribution system needs to be coordinated vertically to fulfill the request for a change in active power flow at the PCC with the transmission system. The chosen cascaded structure of OFO controllers matches the hierarchical structure of the electrical grid and its system boundaries. 

\subsection{Cascaded OFO Controllers}
We define a directed acyclic graph $G=(V,E)$ representing the hierarchy of the controllers, where each of the vertices $V$ represents an individual OFO controller. Edges $E$ represent the unidirectional communication paths between the controllers. An OFO controller can therefore only communicate downstream to request flexibility from any subordinate OFO controller but cannot communicate with controllers on the same or any upper level. Each controller can only have one superordinate controller. 

\begin{figure}[tb]
	\centerline{\includegraphics{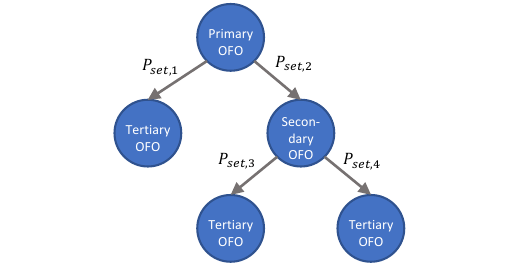}}
	\caption{Directed graph representation of cascaded OFO architecture \cite{zettl2024}}
	\label{fig:hierarchy}
\end{figure}

In total, three different types of OFO controllers can be distinguished. This depends on the location in the control hierarchy and the task:  

\begin{enumerate}
	\item Primary Controller: Only connected to subordinate OFO controllers; supervises top-level grid layer
	\item Secondary Controller: Has sub- and superordinate OFO controllers; provides flexibility to superordinate grid layer
	\item Tertiary Controller: Only has superordinate OFO controllers; provides flexibility to superordinate grid layer
\end{enumerate}
The resulting structure is shown in Fig.~\ref{fig:ofo_simple} and Fig.~\ref{fig:hierarchy}. Unique sets of controllable assets in the grid, as well as observable busses and lines, are assigned to individual controllers. The sets of controllers do not overlap, even if multiple OFO controllers act on the same grid layer, so that each asset only receives set points from one controller. In \cite{klein-helmkamp2024}, the robustness of OFO in case of limited observability and other independent control algorithms in the grid was shown experimentally. Thus, it is not necessary that all DERs and buses are controlled or observed by OFO.

\subsection{Optimization Problem Formulation}
The generic optimization problem is now tailored to the aim of flexibility provision in the form of an active power change at the PCC. Depending on their controller type, as introduced before, and therefore varying positions in the hierarchy and tasks, the optimization problems of individual controllers differ. 
The primary controller is responsible for maintaining safe grid operation regarding voltage and current limits. This should be achieved with minimal changes to the planned dispatch of DERs. $F$ represents the set of devices that can be actuated by an OFO controller. This includes DERs as well as subordinate OFO controllers. $B$ and $N$ are the sets of observed lines and buses. With this, the optimization problem for primary controllers is formulated as in (\ref{eq:opti_prob_primary}).

Note that it is not necessary to include an explicit formulation of the power flow equations, since the physical system solves them implicitly, and the results are integrated via the online measurements. The QP with convex constraints presents a simpler optimization problem in comparison to an AC OPF. A reduction of complexity is further achieved by decoupling the individual problems due to each controller being responsible only for a part of the grid. Measurements at the respective lines and buses provide values for currents and voltages.

\begin{equation} \label{eq:opti_prob_primary}
	\begin{aligned}
		\min_{P_{set},Q_{set}}   \enspace    \Phi = & \sum_{i \in F} ||P_{i,set} - P_{i,planned}||^{2} 
		\\
		\text{s. t.}  \enspace  & V_{min,n}   \leq V_{meas,n}  \leq V_{max,n} &\enspace \forall n \in N\\
		&  I_{min,i}\leq I_{meas,i} \leq I_{max,i} &\enspace \forall i \in B\\
		&  P_{min,j}   \leq P_{set,j}  \leq P_{max,j} &\enspace \forall j \in F\\
		& Q_{min,j}  \leq Q_{set,j} \leq Q_{max,j} &\enspace \forall j \in F 
	\end{aligned}
\end{equation}
While the constraints remain the same as for the primary controller, responding to flexibility request from superordinate controllers is the task of secondary and tertiary OFO controllers. This should be done without violating constraints within their own observability area. The cost function therefore minimizes the difference between the requested flexibility $P_{set}$ and the active power flow $P_{PCC}$ measured at the PCC: 

\begin{equation} \label{eq:opti_prob_secondary}
	\begin{aligned}
		\min_{P_{set},Q_{set}} \enspace \Phi = & ||P_{set,PCC} - P_{meas,PCC}||^{2} 
		\\
		\text{s. t.}  \enspace  & V_{min,n}   \leq V_{meas,n}  \leq V_{max,n} &\enspace \forall n \in N\\
		& I_{min,i}  \leq I_{meas,i} \leq I_{max,i} &\enspace \forall i \in B\\
		&  P_{min,j}   \leq P_{set,j} \leq P_{max,j} &\enspace \forall j \in F\\
		&  Q_{min,j} \leq Q_{set,j}  \leq Q_{max,j} &\enspace \forall j \in F 
	\end{aligned}
\end{equation}
\newline
While this work only considers active power flow in the objective function, it is also possible to provide another degree of freedom by letting the controller determine active and reactive power set points. 

The next sections explore the effect of parametrization of an individual controller on the performance of the flexibility provision. A first simulative case study is carried out on a three-level test system in section \ref{sec:case1} (see also \cite{zettl2024}). In \ref{sec:case2}, a second case study is carried out on a two-level system containing two OFO controllers acting on the same grid layer and receiving set points from the same superordinate OFO controller. Case study 3 in Section~\ref{sec:case3} explores tuning in different grid layers and under adverse conditions. 

\section{Case Study 1: Three-level Test System}
\label{sec:case1}
To show how different choices for the controller gains affect the flexibility provision with the cascaded control structure, a minimal three-level test system is considered in this section.
Table~\ref{tab:test_system} displays the DERs and loads connected to the different voltage levels (low, medium and high voltage) of the test system. The controller cascade consists of three OFO controllers of secondary and tertiary types, each assigned to one of the grid layers. In this work, the primary controller is not modeled explicitly but is assumed to request a fixed operating point of $P_{PCC}=-120\,MW$ at the PCC between the EHV and HV layer. A stationary load flow calculation is performed, hence assuming a constant operating point of the grid that only changes through actions of the OFO controllers. Robustness of the approach has been shown, e.g., in \cite{klein-helmkamp2023}; thus, external disturbances are neglected for simplicity in case studies 1 and 2 and explicitly investigated in case study 3.
 
\begin{table}[h]
	\begin{center}
		\caption{Controller Cascade and Test System for Case Study 1}
		\begin{tabular}{||l c c c c||} 
			\hline
			Controller & \parbox{1cm}{\centering Type} & \parbox{1cm}{\centering Voltage Level} & \parbox{1.5cm}{\centering Controllable DER [MW] }& \parbox{1.5cm}{\centering Controllable Load [MW]}\\ 
			[1ex] 
			\hline\hline
			$OFO_{1}$ & Secondary & HV & 45 & 190 \\ 
			\hline
			$OFO_{2}$ & Secondary & MV & 6.75 & 8.4 \\ 
			\hline
			$OFO_{3}$ & Tertiary & LV & 0.062 & 0.21 \\
			\hline
		\end{tabular}
		\label{tab:test_system}
	\end{center}
\end{table}

\subsection{Uniform Choice of Controller Gain $\alpha$}
\label{sec:uniform}
\begin{figure}[tb]
	\centerline{\includegraphics{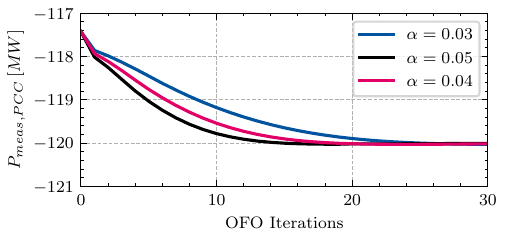}}
	\caption{Influence of controller gain on active power flow at the EHV/HV PCC \cite{zettl2024}}
	\label{fig:fixedalpha}
\end{figure}

The active power flow measured at the PCC converges stepwise to a requested set point if controlled with an OFO controller. This is due to the iterative nature of the approach. The controller gain highly influences the speed of this convergence since in each iteration, the change to the set point vector $u$ is scaled with $\alpha$ (see (\ref{eq:nextinput})). Figure~\ref{fig:fixedalpha} depicts the active power flow controlled by $OFO_1$ at the PCC between the HV and EHV layer. In a first study, every controller is tuned equally to show the dependency between convergence speed and choice of gain $\alpha$. With higher gain, faster convergence to set point $P_{set} = -120\,MW$ is achieved. With $\alpha = 0.05$, the requested set point is reached after 15 iterations. With $\alpha = 0.03$, this is not achieved until after 24 iterations. Thus, an increase in controller gain can significantly reduce the time needed for flexibility provision to the upper grid layer. Higher values can have negative effects as well and lead to unwanted oscillations, as shown in Section~\ref{sec:instability}. 

\subsection{Set Point Tracking}
\begin{figure}[tb]
	\centerline{\includegraphics{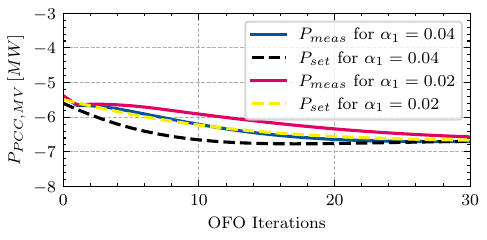}}
	\caption{Active power flow to the requested set point from superordinate OFO controller \cite{zettl2024}}
	\label{fig:setpointtracking}
\end{figure}

Now only the gain of $OFO_1$ is changed while the parameters of the controllers $OFO_2$ and $OFO_3$ are kept constant. Fig.~\ref{fig:setpointtracking} shows the active power flows at the HV/MV PCC for gains of $\alpha_1=0.04$ and $\alpha_1 = 0.02$, while for the other two controllers $\alpha_2 = \alpha_3 = 0.025$ are not varied. The set point that is requested by $OFO_1$ from $OFO_2$ is represented by dashed lines. The flexibility request changes after each iteration before eventually reaching steady state. The measured power flow controlled by $OFO_2$ tracks this changing set point from the superordinate controller until it eventually reaches steady state as well. The speed of flexibility provision therefore depends highly on the parameter choice of the superordinate controller and not only on the subordinate one.

After every iteration, the new set points calculated by a controller are sent to all subordinate controllers and controlled DERs. The iterative nature of this influences how the cascaded structure behaves. By choosing a smaller or larger gain, the convergence speed to steady state and calculated set points can be influenced. 
Each new set point from a superordinate controller changes the optimization problem of a controller; thus, actions taken by secondary and tertiary controllers are directly dependent on  changing set points from superordinate controllers and their respective gains. Following the notion of time scale separation, one would assume that $\alpha_{LV} >> \alpha_{MV} >> \alpha_{MV} $ should hold for optimal performance of the cascaded structure . Here, counterintuitively, a smaller gain value for controllers at higher levels in the cascade improves the behavior of the cascade.

\subsection{Influence of Suboptimal Parameterization}
\label{sec:instability}
\begin{figure}[h]
	\centerline{\includegraphics{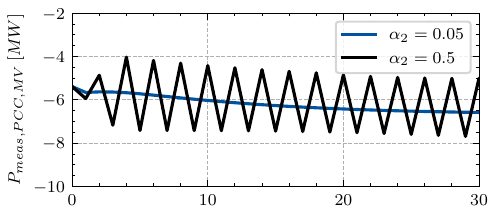}}
	\centerline{\includegraphics{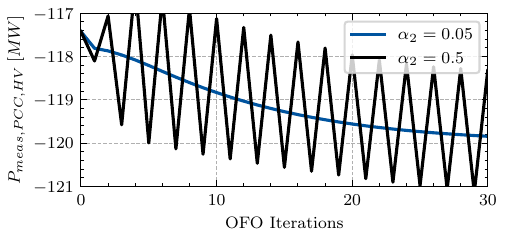}}
	\caption{Active power flow at EHV/HV PCC ($P_{meas,PCC,HV}$) and HV/MV PCC ($P_{meas,PCC,MV}$) depending on choice of $\alpha$ for OFO\textsubscript{2} \cite{zettl2024}}
	\label{fig:instability}
\end{figure}

Fig.~\ref{fig:instability} displays measured active power flow at the HV/MV PCC and at the EHV/HV PCC. The parameters of $OFO_1$ and $OFO_3$ are kept constant at $\alpha_1=\alpha_3=0.02$.  The gain of $OFO_2$ (controlling the MV layer) is varied between $\alpha_2=0.05$ (blue graph) and $\alpha_2=0.5$ (black graph). For $\alpha_2=0.5$, an oscillating measured power flow can be observed at both PCCs. The OFO controller is not able to reach a steady state  within the observed time frame. This behavior can also be observed if the gain of a single  OFO controller that is not implemented in a cascade is chosen wrongly. Here, the gain of $OFO_2$ is chosen to trigger the oscillations on purpose. The results in Fig.~\ref{fig:instability} show that not only the measured power flow controlled by $OFO_2$ displays oscillations, but also the power flow controlled by the superordinate controller $OFO_1$. Note that the controller gain of $OFO_1$ did not change. Therefore, the accuracy and speed of the flexibility provision are significantly influenced by the tuning of individual controllers in the cascade. The parameter choice therefore has to find the optimum between performance and stability even when time-scale separation is assumed. Hence, tuning a cascaded structure is not a trivial problem. The influence of system parameters will be investigated in Section~\ref{sec:case3}.

\section{Case Study 2: Cascaded Interaction in Large-Scale Systems}
\label{sec:case2}
For the second case study, the Simbench Dataset \cite{simbench} was used to create a two-layer system consisting of a medium voltage layer and two low voltage layers with DERs and loads connected according to Table~\ref{tab:simbenchtable}. The two LV networks are assumed to be controlled by two individual OFO controllers of tertiary type. Both LV networks are connected to the same MV network, which is controlled by a single OFO controller of secondary type. Thus, the controllers acting at the LV level are expected to provide the requested active power flow at the respective PCC. The configuration of the cascade can be seen in Table~\ref{tab:cascade2}.

\begin{table}[h]
	\begin{center}
		\caption{Benchmark System for Case Study 2}
		\begin{tabular}{||l c c c c||} 
			\hline
			Subnet & \parbox{1cm}{\centering Nodes} & \parbox{1cm}{\centering Lines} & \parbox{1.5cm}{\centering No. of DER}& \parbox{1.5cm}{\centering No. of Loads}\\ 
			[1ex] 
			\hline\hline
			$MV$ & 119 & 126 & 121 & 120 \\ 
			\hline
			$LV1$ & 14 & 13 & 8 & 28 \\ 
			\hline
			$LV2$ & 110 & 109 & 15 & 129 \\
			\hline
		\end{tabular}
		\label{tab:simbenchtable}
	\end{center}
\end{table}

\begin{table}[h]
	\begin{center}
		\caption{Controller Cascade for Case Study 2}
		\begin{tabular}{||l c c c c||} 
			\hline
			Controller & \parbox{1cm}{\centering Type} & \parbox{1cm}{\centering Voltage Level} & \parbox{1.5cm}{\centering Controllable DER [MW] }& \parbox{1.5cm}{\centering Controllable Load [MW]}\\ 
			[1ex] 
			\hline\hline
			$DSO_{1}$ & Secondary & MV & 62.27 & 47.11 \\ 
			\hline
			$DSO_{2}$ & Tertiary & LV1 & 0.52 & 0.199 \\ 
			\hline
			$DSO_{3}$ & Tertiary & LV2 & 0.2 & 0.61 \\
			\hline
		\end{tabular}
		\label{tab:cascade2}
	\end{center}
\end{table}
\subsection{Influence of Superordinate Controller}
\label{sec:influencesuperordinate}
\begin{figure}[h]
	\centerline{\includegraphics{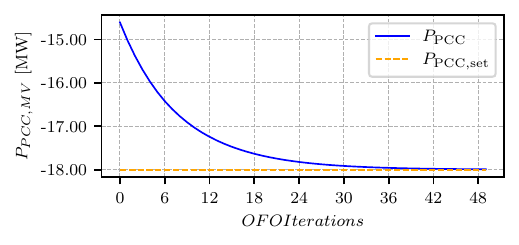}}
	\centerline{\includegraphics{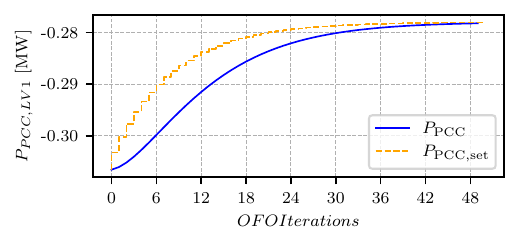}}
	\caption{Active power flow at PCC between HV/MV ($P_{meas,PCC,MV}$) and MV/LV1 ($P_{meas,PCC,LV1}$) grid and respective set point for $\alpha_1=0.001$ and $\alpha_{2,3}=0.1$}
	\label{fig:dso1dso2_stable}
\end{figure}

For a first simulation, the parameters of the controllers were chosen as $\alpha_1=0.001$ for the superordinate controller $DSO_1$ and $\alpha_{2,3}=0.01$ for the subordinate controllers $DSO_2$ and $DSO_3$. The resulting power flows at the PCC between HV and MV layer and the MV and LV layer (only LV1) are shown in Fig.~\ref{fig:dso1dso2_stable}. The set point requested from the superordinate controller is assumed to remain constant over time at $P_{PCC,MV,set} = -18\,MW$. As can be seen in the upper part of the figure, the measured active power flow at the HV/MV PCC converges to the requested set point and reaches a steady state after 40 iterations. In the lower part of the figure, the measured active power flow at the MV/LV1 PCC is depicted, as well as the active power set point requested from controller $DSO_1$. The step wise increase of the requested set point can be noted, changing after each control step until the superordinate controller reaches a steady state. The measured active power flow, which is controlled by the controller $DSO_2$, converges to the requested set point, thus reaching a steady state as well 5 iterations after the superordinate controller. For LV2 controlled by $DSO_3$, a similar behavior is observed. The interdependence of convergence speed between the different control layers, which was already visible in the smaller test system in case study 1, can therefore be confirmed with the larger benchmark system.  

\begin{figure}[t]
	\centerline{\includegraphics{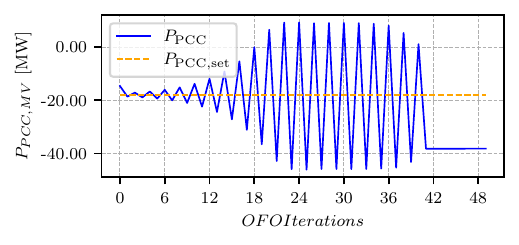}}
	\centerline{\includegraphics{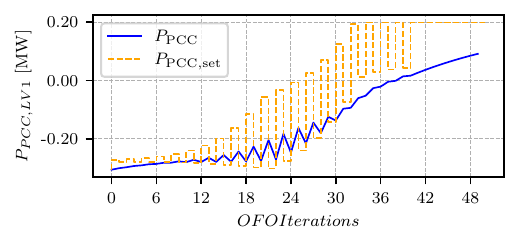}}
	\caption{Active power flow at PCC between HV/MV ($P_{meas,PCC,MV}$) and MV/LV1 ($P_{meas,PCC,LV1}$) grid and respective set point for $\alpha_1=0.01$ and $\alpha_{2,3}=0.1$}
	\label{fig:dso1dso2_unstable}
\end{figure}

\begin{figure}[b]
	\centerline{\includegraphics{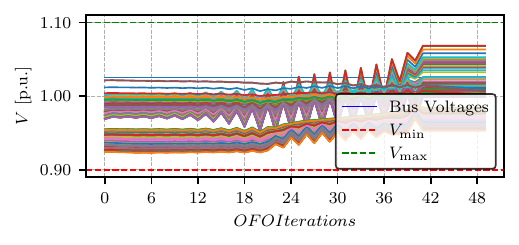}}
	\caption{Bus voltages for $\alpha_1=0.01$ and $\alpha_{2,3}=0.1$}
	\label{fig:busvoltages}
\end{figure}

For the second simulation, the gain of the superordinate controller at MV level, $DSO_1$, was increased to $\alpha_1=0.01$, while the gain of $DSO_2$ and $DSO_3$ was kept at $\alpha_{2,3}=0.01$. The resulting active power flows at the PCC between HV and MV system, as well as MV and LV1, are shown in Fig.~\ref{fig:dso1dso2_unstable}. The controller shows oscillating behavior, similar to what was already shown in Fig.~\ref{fig:instability}. The controller is unable to drive the measured active power flow $P_{PCC,MV}$ to a steady state until no feasible solution to the optimization problem is found in iteration 41. After this, no valid results are produced by the control algorithm. In the lower part of the figure, the active power set point requested from $DSO_2$ is shown. It can be seen that the subordinate controller tracks the oscillations; therefore, it also does not reach a steady state during the simulation. It can be noted that, in comparison to the oscillations produced by the superordinate controller, controller $DSO_2$ reduces the oscillations at LV level significantly after iteration 30.  Fig.~\ref{fig:busvoltages} shows the bus voltages of the whole system (MV and LV layers), where the oscillations are also visible.  

As a summarizing remark, it can be said that attentive tuning of all controllers is important for the operation of the entire system. A smaller gain for the superordinate controller and thus a later arrival at steady state leads to slower provision of flexibility from subordinate grid layers. A higher gain for the superordinate controller allows for larger control steps but can lead to oscillations, as it reduces the appropriate separation of time-scales between the grid layers. With the total installed power in the upper grid layers being higher, more power is actuated in total during one control step, thus possibly leading to oscillations with higher gains. 

\section{Case Study 3: Influence of System Parameters}
\label{sec:case3}

In the third case study, the utilization of OFO in systems with different parameters is evaluated. For this, three grid models from the Simbench data set are selected, according to Table~\ref{tab:simbenchtable2}. Each grid is controlled by one OFO controller. Neither the grid models, nor the controllers are connected in this case study. Each controller is requested to reduce the active power flow at the PCC to the superordinate grid layer by 50\%. 

\subsection{Influence of System Configuration}
\label{sec:systeminfluence}
\begin{figure}[b]	\centerline{\includegraphics{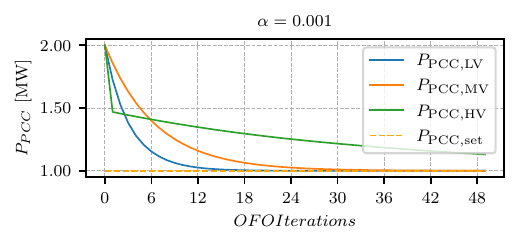}}
	\caption{Active power flow at PCC for different voltage levels and $\alpha = 0.001$}
	\label{fig:comp_voltage_levels}
\end{figure}

Fig.~\ref{fig:comp_voltage_levels} shows the resulting power flow for a gain of $\alpha=0.001$ for all three voltage levels. To establish comparability, the set point is chosen to reduce the active power flow in all networks by half. As can be seen, the convergence speed depends not only on the gain, as shown before, but also on external factors such as system parameters. The figure shows a higher convergence speed for the OFO controller governing the LV grid as for the OFO controller responsible for the HV grid. While the set point in the LV grid is reached after approximately 18 iterations, the set point in the HV grid is not reached within the 49 iterations that are displayed. The results suggest increasing the gain for networks on higher voltage levels. The selected HV grid model features a higher complexity regarding its topology and its system states are closer to operational limits than those of the LV grid, thus potentially preventing fast convergence. 

\begin{table}[h]
	\begin{center}
		\caption{Benchmark Systems}
		\begin{tabular}{||l c c c c||} 
			\hline
			Voltage Level & \parbox{1cm}{\centering Nodes} & \parbox{1cm}{\centering Lines} & \parbox{1.5cm}{\centering No. of DER}& \parbox{1.5cm}{\centering No. of Loads}\\ 
			[1ex] 
			\hline\hline
			$LV$ & 111 & 109 & 15 & 129 \\ 
			\hline
			$MV$ & 136 & 147 & 134 & 139 \\ 
			\hline
			$HV$ & 114 & 145 & 124 & 58 \\
			\hline
		\end{tabular}
		\label{tab:simbenchtable2}
	\end{center}
\end{table}
% ToDo: Anpassen!

\subsection{Influence of Adverse Conditions}
\label{sec:adverseconditions}
The robustness of OFO against model mismatch and measurement errors has been shown before, as discussed in Section~\ref{sec:relatedwork}. However, the influence of tuning OFO controllers for set-point tracking considering non-ideal information in power system operation has not been investigated. This section gives insights into the influence of controller gain on convergence speed considering different uncertainties. The same LV grid as in the previous section, with parameters listed in Table~\ref{tab:simbenchtable2}, is simulated and a set point of $P_{PCC,set}=0W $ is applied.
To account for measurement uncertainty, the vector of measured states is modeled as subject to additive Gaussian noise. Let $x$ denote the full system state and let $y$ represent the available measurements, which may correspond to a subset of $x$. The measurements satisfy
\begin{equation}
    y = x + v,
\end{equation}
where $v \sim \mathcal{N}(0, R)$ is a zero-mean normal random vector with covariance matrix $R$. The noise is assumed to be independent across time.
To account for model mismatch, the sensitivity matrix is altered in a similar way. Let $h_{i,j}^{true}$ be a correct entry of the sensitivity matrix. The imperfect sensitivity matrix then satisfies
\begin{equation}
    h_{i,j} = h_{i,j}^{true} \cdot \eta 
\end{equation}
where $\eta \sim \mathcal{U}(1-b,1+b)$ follows a uniform distribution. It is assumed that the mismatch of physically close elements is correlated. 

\begin{figure}[t]
    \centering
    \includegraphics[width=\linewidth]{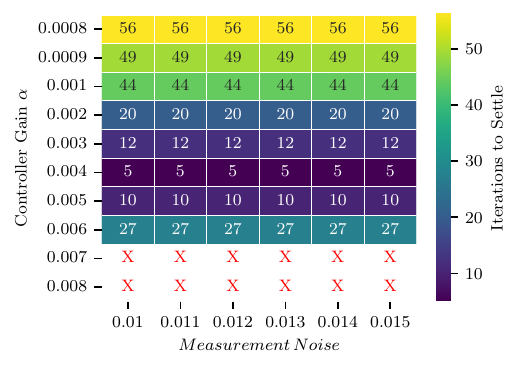}
    \caption{Effects of measurement noise and gain on convergence}
    \label{fig:alpha_noise_matrix}
\end{figure}

Figure~\ref{fig:alpha_noise_matrix} shows the number of iterations needed to converge to the requested set point with a tolerance of $\epsilon = 0.01\%$. As can be seen, both a too high gain, e.g. $\alpha=0.007$ or $\alpha=0.008$ prevent convergence entirely. While for the lower range an increase in $\alpha$ does reduce the number of iterations to a minimum of 5 iterations for $\alpha=0.004$, a further increase for higher gains leads to a slower convergence again. For the simulated case, the measurement noise does not influence the number of iterations to settle. Therefore, for the simulated case, measurement noise does not need to be considered during the tuning process.  

\begin{figure}[t]
    \centering
    \includegraphics[width=\linewidth]{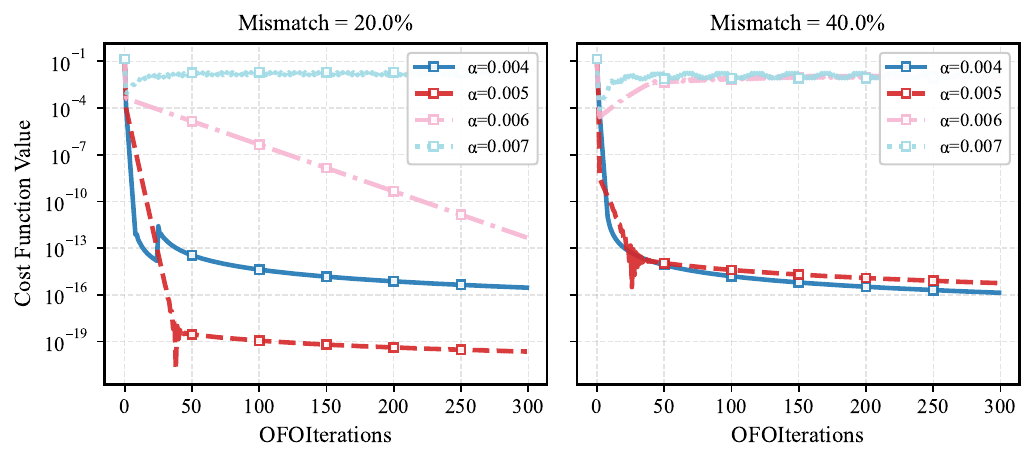}
    \caption{Effects of model mismatch and gain on convergence}
    \label{fig:alpha_mismatch}
\end{figure}

Fig.~\ref{fig:alpha_mismatch} shows the cost function value for 300 iterations of the OFO controller with applied model mismatch of $b=0.2$ and $b=0.4$ for different values of $\alpha$. Here, the measurements are assumed to be noiseless. As can be seen, for the highest simulated gain of $\alpha=0.07$, no decrease in the cost function value is achieved, and the trajectory is oscillating, in both cases. A higher model mismatch increases the oscillating behavior. For a small gain of $\alpha=0.004$, the trajectory remains almost unchanged, while for $\alpha=0.005$ and $\alpha=0.006$, the cost function value increases with increasing model mismatch. This indicates a reduced ability to track the requested set point and thus a reduced performance of the controller. Therefore, smaller gain values are recommended if higher model mismatch is to be expected.  

\section{Conclusion}
In this work, a cascaded structure of online feedback optimization controllers was implemented and the influence of individual controller tuning was investigated.
Each OFO controller in the structure was assigned to a specific grid layer with a set of controllable DERs. For the algorithm, an optimization problem is integrated in closed loop with the physical system, thus forming a feedback controller. Closing the loop and using online measurements provides robustness and allows fast provision of flexibility while ensuring grid constraints. The objective function is formulated with the aim of changing the power flow at the PCC to the superordinate grid layer to provide flexibility.  The implemented cascade is first studied on a three-level distribution system consisting of low to high voltage levels and then confirmed on a larger two-level benchmark system containing medium and low voltage layers. The approach shows high suitability for vertical coordination of flexibility. 

An attentive tuning of individual controllers in the cascade is crucial for the flexibility provision, as a high interdependency between the tuning of individual controllers and the behavior of other controllers could be observed. A smaller gain and thus a slower convergence speed of superordinate controllers reduces convergence speed of subordinate controllers. It is therefore recommended to increase the gain of controllers in higher voltage levels in comparison to the ones in lower voltage levels. Oscillations caused by too high gains in one layer can lead to oscillations in other layers as well. Thus, the trade-off between stable behavior on one hand and fast flexibility provision on the other hand has to be considered. While the introduction of measurement noise shows little effects, the presence of model mismatch influences the parametrization. Smaller gains allow for convergence even with higher model mismatch. The approach in this work can be further investigated in future research. Aspects to be considered can include the optimal choice of cycle times, further optimizing controller interactions or effects of faulty communication between the controllers.

\end{document}